\documentstyle[11pt]{article}
\begin{document}
\pagestyle{plain}

\title{\huge \bf Radiation of the  blackbody in the external field}
\large
\author{Miroslav Pardy\\[7mm]
%Institute of Plasma Physics ASCR\\
%Prague Asterix Laser System, PALS\\
%Za Slovankou 3, 182 21 Prague 8, Czech Republic\\
%and\\
Department of Physical Electronics \\
and\\
Laboratory of Plasma physics\\[5mm]
Masaryk University \\
Kotl\'{a}\v{r}sk\'{a} 2, 611 37 Brno, Czech Republic\\
e-mail:pamir@physics.muni.cz}
\date{\today}
\maketitle

\vspace{10mm}
\baselineskip 15pt
\begin{abstract}
The blackbody is considered in the external general field. 
The additional coefficients of stimulated emission and
stimulated absorption are introduced into the Einstein mechanism and the
generalized Planck formula is derived. The Einstein and Debye formula 
for the specific heat is  possible to generalize.
The Bose-Einstein statistics is broken in the external field. 
The relation of the theory  to the sonoluminescence, the 
relic radiation and solar spectrum is considered.
\end{abstract}

\vspace{3mm}


\vspace{13mm}

\section{Introduction}

The distribution law of photons inside of the so called blackbody was
derived in 1900 by Planck (1900, 1901), (Sch\"opf, 1978). 
The derivation was based on the
investigation of the statistics of the system of oscillators. 
Later Einstein (1917)  derived the Planck formula from the Bohr model of atom. 
Bohr created two postulates
which define the model of atom: 1. every atom can exist in the discrete
series of states in which electrons do not radiate even if they are moving at
acceleration  (the postulate of the stationary states), 2. transiting electron
from the one stationary state to other, emits the energy according to
the law

$$\hbar\omega = E_{m} - E_{n},\eqno(1)$$
where $E_{m}$ is the energy of an electron in the
initial state, and $E_{n}$ is the energy of the final state of an
electron  to which the transition is made and $E_{m} > E_{n}$.

Einstein introduced coefficients of spontaneous  and stimulated emission
$A_{mn}, B_{mn}, B_{nm} $. In the case of spontaneous emission,
the excited atomic state decays without external stimulus as an analogue
of the natural radioactivity decay. The energy of the emitted photon
is given by the Bohr formula (1). In the process of the stimulated
emission the atom is induced by the external stimulus to make the
same transition. The external stimulus is a blackbody photon that has
an energy given by the Bohr formula (1).

If the number of the excited atoms is equal to $N_{m}$, the emission
energy per unit time  conditioned by the spontaneous transition from
energy level ${E_{m}}$ to energy level ${E_{m}}$ is

$$P_{spont. \; emiss.} = N_{m}A_{mn} \hbar\omega, \eqno(2) $$
where $A_{mn}$ is the coefficient of the spontaneous emission.

In case of the stimulated emission, the coefficient  $B_{mn}$
 corresponds to the transition of an electron
 from energy level ${E_{m}}$ to energy level ${E_{n}}$ and coefficient
 $B_{nm}$ corresponds to the transition of an electron
 from energy level ${E_{n}}$ to energy level ${E_{m}}$. So, for the
 energy of the stimulated emission per unit time we have two formulas :

$$P_{stimul. \; emiss.} = \varrho_{\omega}N_{m}B_{mn} \hbar\omega \eqno(3) $$

$$P_{stimul. \; absorption} = \varrho_{\omega}N_{n}B_{nm}
\hbar\omega. \eqno(4) $$

If the blackbody is in thermal equilibrium, then the number of
transitions from ${E_{m}}$ to ${E_{n}}$ is the same as 
from ${E_{n}}$ to ${E_{m}}$ and we write:

$$N_{m}A_{mn}\hbar\omega + N_{m}\varrho_{\omega}B_{mn}\hbar\omega =
N_{n}\varrho_{\omega}B_{nm}\hbar\omega,  \eqno(5)$$
where $\varrho_{\omega}$ is the density of the photon energy of the blackbody. 

Then, using the Maxwell statistics

$$N_{n} = De^{-\frac{E_{n}}{kT}},\quad  N_{m} = De^{-\frac{E_{m}}{kT}},
\eqno(6)$$
we get:

$$\varrho_{\omega} = \frac{\frac{A_{mn}}{B_{mn}}}
{\frac{B_{nm}}{B_{mn}}e^{\frac{\hbar\omega}{kT}} - 1}.\eqno(7)$$

The spectral distribution of the blackbody does not depend on the
specific atomic composition of the blackbody and it means the formula
(7) must be so called the Planck formula:

$$\varrho_{\omega} = \frac{\hbar\omega^3}{\pi^2 c^3}\frac{1}
{e^{\frac{\hbar\omega}{kT}} - 1}.\eqno(8)$$

After comparison of eq. (7) with eq. (8) we get:

$$B_{mn} = B_{nm} = \frac{\pi^2 c^3}{\hbar \omega^3}A_{mn}. \eqno(9)$$

It means that the probabilities
of the stimulated transitions from ${E_{m}}$ to ${E_{n}}$ 
and from ${E_{n}}$ to ${E_{m}}$ are
proportional to the probability of the spontaneous transition
$A_{mn}$. So, it is sufficient to determine only one of the
coefficient in the description of the radiation of atoms.
 
The Planck law (8) can be also written as 

$$\varrho_{\omega} = G(\omega)<E_{\omega}> =
\frac{\omega^2}{\pi^2 c^3}\frac{\hbar\omega}
{e^{\frac{\hbar\omega}{kT}} - 1}.\eqno(10)$$
where the term 
$$<E_{\omega}>  = \frac{\hbar\omega} {e^{\frac{\hbar\omega}{kT}} - 1}
\eqno(11)  $$  
is the average energy of photons in the blackbody and 

$$G(\omega) = \frac{\omega^2}{\pi^2 c^3}\eqno(12)$$
is the number of electromagnetic modes in the interval $\omega, \omega
+ d\omega$. 

Let us remark that coefficients $A_{mn}$ of the so called spontaneous
emission cannot be specified in the framework of the classical 
thermodynamics, or, statistical physics. They  can be determined 
only by the methods of quantum
electrodynamics as the consequences of the so called radiative
corrections. So, the radiative corrections are hidden external stimulus, 
which explains the spontaneous emission.

\section{N-dimensional blackbody}

The problem of the N-dimensional blackbody is related to the 
dimensionality of space, or, of space-time and some ideas on the
dimensionality was also
expressed and analyzed in recent time by author (Pardy, 2005). 

The experimental
facts following from QED experiments, galaxy formation and formation of the
molecules DNA, prove that the external space is
3-dimensional. With regard to the Russell philosophy of mathematics, there
is no possibility to prove the dimensionality of space, or, space-time
by means of the pure mathematics, because the statements of mathematics
are nonexistential. The existence of the external world cannot be also
proved by pure mathematics. However, if there is an axiomatical system
related to the external world and reflecting correctly the external world, 
then, it is possible to do many
predictions on the external world by pure logic. 
This is the substance of exact sciences. We know for instance that the
success of special theory of relativity is is based on the adequate
axiomatical system and on logic. 

In case of the n-dimensional blackbody, the number of
modes can be determined (Al-Jaber, 2003). We use here alternative 
and elementary derivation.
We know, that in case that the electromagnetic field is in a box 
of the volume  $L^{n}$, the wave vector ${\bf k}$ is
quantized and the elementary volume in the k-space is
$\Delta_{0n} = (2\pi)^{n}/L^{n}$.  

The elementary volume of the n-dimensional k-space is evidently
the volume $d V_{n}$ between spheres with radius $k$ and $k + dk$:

$$dV_{n} = d\left\{\frac {2  \pi^{n/2}}{n\Gamma\left(\frac {n}{2}\right)}
k^{n}\right\} = 
\frac {2\pi^{n/2}}{\Gamma\left(\frac {n}{2}\right)}
k^{n-1}dk, \eqno(13)$$
where $\Gamma(n)$ is so called Euler gamma-function defined in the internet mathematics  (http://mathworld.wolfram.com/GammaFunction.html) as 

$$\Gamma(x) = \int_{0}^{\infty}e^{-t}t^{x-1}dt; \quad 
\Gamma(n/2) =  \frac{(n-2)!!\sqrt{\pi}}{2^{(n-1)/2}}.\eqno(14)$$

The number of electromagnetic modes involved inside the spheres between $k$
and $k + dk$ is then, with
$\omega = ck$,  or $k = \omega/c$ and $dk = d\omega/c$,

$$2\times \frac{dV_{n}}{\Delta_{0n}} = 2\times 
\frac {1}{\Gamma\left(\frac {n}{2}\right)}
\frac {1}{2^{(n-1)}\pi^{n/2}}L^{n}\frac {\omega^{n-1}d\omega}{c^{n}},
\eqno(15)$$
where isolated number 2 expresses the fact that light has 2
polarizations. It means that the number of modes $G_{n}$, in volume
$V$,  involved 
between spheres $k$ and $k + dk$, or, $\omega$ and $\omega + d\omega$, is 

$$G_{n}(\omega)d\omega  = 2\times
\frac {1}{2^{(n-1)}}\frac {1}{\Gamma\left(\frac {n}{2}\right)}
\frac {1}{\pi^{n/2}}L^{n}\frac {\omega^{n-1}}{c^{n}} d\omega.
\eqno(16)$$

For the spectral distribution, we have 

$$\varrho_{n\omega}d\omega = <E_{n}>\frac{G(\omega)d\omega}{L^n} = 
<E_{n}>g(\omega)d\omega, \eqno(17)$$
where the average value of energy $<E_n>$ is given by eq. (11).

The spectral distribution of the n-dimensional blackbody has the
final form following from eq. (17):

$$\varrho_{n\omega} = \frac{\hbar \omega }{e^{\frac{\hbar\omega}{kT}} - 1}\;
\frac {1}{2^{(n-2)}}\frac {1}{\Gamma\left(\frac {n}{2}\right)}
\frac {1}{\pi^{n/2}}\frac {\omega^{n-1}}{c^{n}}.
\eqno(18)$$

\section{Blackbody in the external field}

The Einstein derivation does not
consider the situation where the blackbody is influenced by some
external field. Einstein was not motivated by the case of so called 
sonoluminescence, or by relic radiation and so on, which play at present 
time in physics the fundamental role. Now, if we include some external
nonspecified field, it is necessary to introduce transition 
coefficients $C_{mn}, C_{nm},$. From the equilibrium condition

$$N_{m}A_{mn} + N_{m}\varrho_{\omega}B_{mn}  + N_{m}C_{mn}=
N_{n}\varrho_{\omega}B_{nm}  + N_{n}C_{nm}\eqno(19)$$
and Maxwell statistics (6) we get:

$$A_{mn} + \varrho_{\omega}B_{mn}  + C_{mn} =
e^{\frac{\hbar\omega}{kT}}\varrho_{\omega}B_{nm}  + e^{\frac{\hbar\omega}{kT}}
 C_{nm} ,\eqno(20)$$
which can be modified using the definitions

$$B_{nm} = B_{mn} = \frac{\pi^2 c^3}{\hbar \omega^3}A_{mn}; \quad
\frac{C_{mn}}{A_{mn}} = \beta_{mn}; \quad \frac{C_{nm}}{A_{mn}} =
\beta_{nm}\eqno(21)$$
as follows:

$$1 + \beta_{mn} - e^{\frac{\hbar\omega}{kT}}\beta_{nm} = \varrho_{\omega}
\frac{\pi^2 c^3}{\hbar \omega^3}\left(e^{\frac{\hbar\omega}{kT}} - 1
\right). \eqno(22)$$

The generalized Planck law follows from the equation (22):

$$ \varrho_{\omega} = \frac{\hbar\omega^3}{\pi^{2} c^3}\frac{1}
{e^{\frac{\hbar\omega}{kT}} - 1} +
\frac{\hbar\omega^3}{\pi^{2} c^3}\;\frac{P(\omega) -
 Q(\omega)e^{\frac{\hbar\omega}{kT}}}{e^{\frac{\hbar\omega}{kT}} - 1},
\eqno(23) $$
where $P =  \beta_{mn}, Q =  \beta_{nm}$ are some function which
must be calculated using the advanced solid state physics and 
advanced quantum mechanics. In order to get the second term in
eq. (23) finite for $\omega \to \infty$, it is possible to postulate 
the mathematical structure of $Q(\omega)$ as follows: 
$Q(\omega) = R \exp(-S\omega)$, where $R, S$ are some constants. 

\section{Discussion}

The original Planck derivation of the blackbody radiation was based
on the relation between the entropy of the system and the internal
energy of the blackbody denoted by Planck as $U$.

While from the postulation of the relation 

$$\frac{d^2S}{dU^2} = - \frac{const}{U}\eqno(24)$$
the  Wien law follows, the a priory generalization of eq. (24) gives
new physics. The generalization was postulated by Planck in the following form:

$$\frac{d^2S}{dU^2} = - \frac{k}{U(\varepsilon + U)}.\eqno(25)$$

The first integration can be performed  using the integral 

$$\int\frac{dx}{x(a + bx)} = -
\frac{1}{a}\ln\left|\frac{a}{x} + b\right|.\eqno(26)$$

We get the following result

$$\frac{1}{T} =  \frac{dS}{dU} =  \frac{k}{\varepsilon }\ln
\left(\frac{\varepsilon}{U} + 1\right).\eqno(27)$$

The solution of eq. (27) is  

$$U = \frac{\varepsilon}{{\rm e}^{\varepsilon/kT} -1}.\eqno(28)$$ 

The general validity of the Wien law

$$\frac{dS}{dU} = \frac{1}{\nu}f\left(\frac{U}{\nu}\right)\eqno(29)$$
confronted with the equation (27) gives the famous Planck formula 

$$\varepsilon = h\nu.\eqno(30)$$

The next step of Planck was to find the appropriate physical
statistical system which led to the correct power spectrum of the
blackbody. This model was the thermal reservoir of the independent
electromagnetic oscillators with the discrete energies. 

We know that Einstein later identified the reservoir with the
oscillators in the crystalline medium in order to get his famous heat
capacity with the high temperature limit via the Dulong-Petit law.
He supposed that the frequency of the crystalline medium is
$\omega_{E}$, $\omega_{E}$ being the Einstein frequency. 
Then $U = 3N<E>$, N being the number of oscillators. Then, 

$$C_{V} = \frac{\partial U}{\partial T} =
\frac{3Nk\left(\frac{\hbar\omega_{E}}{kT}
\right)^{2}\exp(\hbar\omega_{E}/kT)}{\left(\exp(\hbar\omega_{E}/kT) -
1\right)^{2}}.\eqno(31)$$

For $kT>>\hbar\omega_{E}$ it is $\exp(\hbar\omega_{E}/kT) \approx 1 + 
\hbar\omega_{E}/kT$ 
and we get 

$$C_{V} \approx 3Nk. \eqno(32)$$

For $kT << \hbar\omega_{E}$ we get from the formula $U$ 
the following formula the approximation 

$$U \approx 3N\hbar\omega_{E} e^{-\frac{\hbar\omega_{E}}{kT}},\eqno(33)$$
from which 

$$C_{V} = \frac{\partial U}{\partial T} =  
\frac{3Nk\hbar^{2}\omega_{E}^{2}}{(kT)^{2}}e^{-\frac{\hbar\omega_{E}}{kT}}
\eqno(34)$$
and evidently $C_{V} \rightarrow 0$, when $T \rightarrow 0$, which is in
a agreement with experiment.  

However, the absolute harmony of theory with experiment was not
achieved, because experimentally it is $C_{V} \approx T^{3}$, 
which does not follow from the formula (34).

Later Debye in 1912 generalized the Einstein model in order to get more
realistic formula for the heat capacity of every crystalline matter
(Reissland, 1973).
Debye assumed  that the solid state can be represented by the
continual medium with many modes. However, if we consider a continual medium,
then there are the infinite number of frequencies. Debye resolved this
contradiction by the normalization condition that the number of
frequencies must be 3N. Or, 

$$\int_{0}^{\omega_{D}}G(\omega)d\omega = 3N ,\eqno(35)$$ 
where $G(\omega)$ is the number of phonon modes in the
three-dimensional continuum. Or, 

$$G(\omega)d\omega = \frac{V}{2\pi^2}\left(\frac{2}{v_{t}^{3}} + 
\frac{1}{v_{l}^{3}}\right)\omega^{2}d\omega = 
\frac{3\omega^{2}V}{2\pi^{2}c^{3}}d\omega, \eqno(36)$$
where $v_{t}$ is the transversal velocity with two polarization 
and $v_{l}$ is the longitudinal velocity of sound and we  have defined
$c$ 
by the equation 

$$\frac{3}{c^{3}} = \left(\frac{2}{v_{t}^{3}} + 
\frac{1}{v_{l}^{3}}\right).\eqno(37)$$ 

From equation (35) follows the Debye maximal frequency

$$\omega_{D}^{3} = 18 \pi^{2} \frac{N}{V}\left(\frac{2}{v_{t}^{3}} + 
\frac{1}{v_{l}^{3}}\right)^{-1}.\eqno(38)$$

The average energy is then of the form:

$$U = \frac{3V\hbar}{2\pi^{2}c^{3}} \int_{0}^{\omega_{D}}
\frac{\omega^{3}d\omega}{e^{\frac{\hbar\omega}{kt}}- 1},\eqno(39)$$ 

which can be written using the substitution $x = \hbar\omega/kT$ in
the following form:

$$U = \frac{3V(kT)^{4}}{2\pi^{2}c^{3}\hbar^{3}}
 \int_{0}^{\frac{\hbar\omega_{D}}{kT}}
\frac{{x^{3}}dx}{e^{x}- 1}.\eqno(40)$$ 

The natural approximation is for $kT << \hbar\omega_{D}$ where  the
upper limit of the integral is infinity and we have with 

$$\int_{0}^{\infty}
\frac{{x^{3}}dx}{e^{x}- 1} = \frac{\pi^{4}}{15}\eqno(41)$$

$$U = \alpha T^{4}; \quad  
\alpha = \frac{\pi^{2}k^{4}}{10 c^{3}\hbar^{3}}V. \eqno(42)$$

The corresponding specific heat in the last approximation is $C_{V} = 
\partial U/\partial T = 4 \alpha T^{3}$.

It can be easily to show that if we introduce the debye temperature by
the relation 

$$T_{D} = \frac{\hbar\omega_{D}}{k},\quad  \omega_{D} = 
c\left(\frac{6\pi^{2}N}{V}\right)^{1/3}\eqno(43)$$  
then 

$$U = 3NkT D\left(\frac{T_{c}}{T}\right); \quad D(y)= \frac{3}{y^3}
\int_{0}^{y}\frac{x^{3}dx}{e^{x} - 1}.\eqno(44)$$  

The corresponding heat capacity is then given by the formula

$$C_{V} = 3Nk\left\{D\left(\frac{T_{c}}{T}\right) - \frac{T_{c}}{T}
D'\left(\frac{T_{c}}{T}\right)\right\}.\eqno(45)$$ 
 
The Debye formula is only an approximation
because the frequency spectrum of the crystalline and noncrystalline
medium is very complex. There are also many defects in the
crystalline matter and it means that the determination of this
spectrum is not easy mathematical problem. 
Some theory of the determination of such
spectrum was discussed for instance by Maradudin (1966) and others. 

Nevertheless, after some elementary approximation, Debye was able to
derive the heat capacity of his model of crystal in the following form
(Rohlf, 1994):

$$C_{V} = \frac{12 \pi^4}{5}Nk \left(\frac{T}{T_{D}}\right)^{3};
\quad T_{D} = \hbar\omega_{D}/k .\eqno(46)$$

Of course, the Debye model is in the better agreement with experiment
that the Einstein model.

The formula (23) can be easily written in the form 

$$ \varrho_{\omega} = \frac{\omega^{2}}{\pi^{2} c^3}\frac{\hbar\omega}
{e^{\frac{\hbar\omega}{kT}} - 1}(1 +
P(\omega) -  Q(\omega)e^{\frac{\hbar\omega}{kT}}),\eqno(47) $$
which means that the spectrum of the electromagnetical modes 
inside of the blackbody is modified. If we want to involve the
external field in the Einstein model, Debye model and 
more sophisticated models, then the formal operation 
consists in the following elementary transformation

$$\frac{1}{e^{\frac{\hbar\omega}{kT}} - 1} \longrightarrow 
\frac{1}{e^{\frac{\hbar\omega}{kT}} - 1}
\left(1 + P(\omega) -  Q(\omega)e^{\frac{\hbar\omega}{kT}}\right)
\eqno(48) $$

If this formal procedure has the physical meaning, then 
it is the generalization of the Bose-Einstein statistics (BE). 
The known generalization of the BE statistics and Fermi-Dirac (FD) 
statistics is
known as the parastatistics. It is defined as the statistics where in
one energetical state can be $p$ particles. If $p$ is arbitrary, we
get the BE statistics and if $p = 1$, we get so called FD statistics.
The derivation of parastatistics is involved in physics as an
exercise of combinatorics (Isihara, 1971).

Here, we consider the statistics which is generated by
dynamics. The historical analogue of this approach 
is the Maxwell statistics based on
the dynamics of particles in reservoir and Einstein statistics (BE) of
photons inside of  the blackbody, derived on the basis of 
the dynamics of emission and absorption of photons. 
Our derivation is also based on the dynamics of emission and 
absorption of photons with the additional terms.

The different generalization of the  Bose-Einstein  statistics was realized
for instance by Bekenstein (Bekenstein et al., 1994) in order to explain so
called the gray-body radiation. 

However, the  idea of modification of the Bose-Einstein statistics or
is forbidden by the quantum field theory where there is the strong
connection between spin and statistics. The particles with the integer
spins are bosons and the particles with the half-integer spins are
fermions (Berestetzkii et al.,
1999). Only if we accept the idea that the statistics of photons and
phonons is of the dynamical origin, then we can also write the generalized
heat capacity of solid state.

The next possibility to modify the statistics 
is to write the Planck hypothesis (25) in the generalized form 

$$\frac{d^{2}S}{dU^{2}} = \frac{\alpha}{\sum_{n}a_{n}U^{n}}\eqno(49)$$
and then to determine the modified Planck law for the specific thermal
systems. To our knowledge this way was not still realized.

The formula (23) can be also related to the solar spectrum which is
according to the NASA expertise not absolutely identical with the
spectrum of the blackbody (Thekaekara, 1994). 

The formulas for the modified specific heat  
can be eventually considered in case of the investigation of the
specific heat of the carbon nanostructures in some nonspecified
external field. Some investigation was performed by Li and Chou
(2005). The theory and experiment with the carbon
nanonstructures is at present time very actual area of investigation
and it means that every new information of the physical properties of
such structures is useful.

The formula (23) can be probably related to the spectrum of the
sonoluminescence, because the external field is the
ultrasound. Sonoluminescence is defined as 
the emission of short burst of light
from imploding bubbles in a liquid when excited by ultrasound. 
The effect of sonoluminescence was discovered at the University of
Cologne in 1934. Franzel and Schultes  put an ultrasound transducer in
a tank  of photographic developer fluid. They hoped to speed up the
development process. Instead, they noticed tiny dots on the film
after developing, and realized that the bubbles in the fluid
were emitting light with the ultrasound turned on. The bubbles are
very small when they emit the light. About 1 micrometer in diameter.
The high compression of a small bubble of fluid is similar to the
explosive compression of a pellet of material by laser beams, one of
the methods proposed for the nuclear fusion. 

The situation in cosmology is a such the ``cosmic
microwave background'' (CMB) discovered in 1965 by Penzias and Wilson is 
of the Planck distribution of the temperature 2.7 Kelvin. 
This radiation was predicted by Gamow as a consequence of the Big Bang.   
By the early 1970' it became clear that the CMB sky is of the dipole
form (it is hotter in one
direction and cooler in the opposite direction with the temperature
difference being a few mili Kelvin). The dipole form can be explained
by the motion our galaxy with regard to the rest of universe. However,
at some level one expects to see irregularities, or anisotropies of
CMB, and then it is not excluded that the generalized Planck formula
will be appropriate for the description of CMB.
 
\vspace{15mm}

{\bf References}

\vspace{5mm}
\noindent
Al-Jaber, Sami M. (2003). Planck's spectral distribution law in N dimensions,
{\it Int. Journal of Theor. Phys.} {\bf 42}, No. 1, 111.\\[2mm]
Bekenstein, J. D. (1994).
Universality in gray-body radiance: Extending Kirchhoff's
law to the statistics of quanta, Phys. Rev. Lett. {\bf 72}, 
No. 16, 2512.\\[2mm]
Berestetzkii, V. B., Lifshitz, E. M. and Pitaevskii, L. P. (1999).
Quantum electrodynamics, (Butterworth-Heinemann, Oxford).\\[2mm]
Einstein, A. (1917). Zur quantentheorie der Strahlung, {\it Physikalische
Zeitschrift}, {\bf 18}, 121.\\[2mm]
Isihara, A. (1971). Statistical mechanics, Academic Press, New
York--London. \\[2mm]
Li C. and Chou T-W. (2005). Modelling of heat capacities of
multiwalled carbon nanotubes by molecular structural mechanics,
Materials Science and Engineeering A 409, 140-144.\\[2mm]
http://mathworld.wolfram.com/GammaFunction.html\\[2mm]
Maradudin, A. A. (1966). Theoretical and experimental aspects of the
effects of point defects and disorder on the vibrations of crystals,
(Academic press Inc., New York - London).\\[2mm]
Pardy, M. (205). Light in metric space-time and its deflection by the 
screw dislocation, Horizons in World Physics, Volume 248, 
ed. Reimer, A. (Nova Science Publishers, Inc.).\\[2mm]
Planck, M. (1900). Zur Theorie des Gesetzes der Energieverteilung im
Normalspektrum, {\it Verhandlungen deutsch phys., Ges.}, {\bf 2}, 237.; 
ibid: (1901). {\it Ann. Phys.}, {\bf 4}, 553.\\[2mm]
Reissland, J. A. (1973). The physics of phonons, 
(John Wiley \& Sons Inc., New York)).\\[2mm]
Rohlf, J. W. (1994). Modern physics from $\alpha\; {\rm to}\; Z^{0}$, (John
Wiley \& Sons LTD., London - New York).\\[2mm]
Sch\"opf, H-G. (1978).Theorie der W\"armestrahlung in
historisch-kritischer Darstellung, (Alademie/Verlag, Berlin).\\[2mm]
Thekaekara, M. P. (1970). The Solar constant and Solar spectrum measured
from a research aicraft, NASA technical report R-351 (Reproduced also in
(Rohlf, 1994)).

\end{document}